\documentclass[aps,12pt,final,notitlepage,oneside,onecolumn,nobibnotes,nofootinbib,superscriptaddress,showpacs,centertags,amsmath,amsfont,amssymb,amscd
 ]%
{revtex4}
\def\bb {\begin {eqnarray}}
\def\ee {\end {eqnarray}}

\usepackage[all]{xy}
\usepackage[english]{babel}

\begin{document}\selectlanguage{english}
\title{Contraction based classification of supersymmetric extensions
of kinematical Lie algebras}
\author{\firstname{R.}~\surname{Campoamor-Stursberg}}
\affiliation{IMI, Universidad Complutense de Madrid, Spain}
\author{\firstname{M.}~\surname{Rausch de Traubenberg}}
\affiliation{Universit\'e Louis Pasteur, Strasbourg, France}

\begin{abstract}
We study supersymmetric extensions of classical kinematical
algebras from the point of view of contraction theory. It is shown
that contracting the supersymmetric extension of the anti-de
Sitter algebra leads to a hierarchy similar in structure to the
classical Bacry-L\'evy-Leblond classification.
\end{abstract}

 \maketitle

\section{Introduction}
The contraction approach has been systematically applied in
physics, among other problems, to classify the possible classical
kinematical groups, basing on space isotropy and assuming that
time-reversal and parity are automorphisms of the kinematical
group, as well as non-compactness of one-parameter subgroups
generated by boosts \cite{bl}. Within this frame, all kinematical
models arise as contractions of the de Sitter Lie algebras. It is
therefore natural to ask whether for the supersymmetric extensions
of the latter algebras, constructed in supersymmetric models, a
similar procedure and classification holds, at least for those
extensions proven to be of physical interest.

\medskip

The main objective of this work is to extend the classical
kinematical classification of Bacry and L\'evy-Leblond (BBL
classification) to the supersymmetric case, using generalized
In\"on\"u-Wigner contractions. By means of this procedure, we show
that supersymmetric extensions of kinematical algebras considered
in the literature \cite{MdM,wb} fit into a contraction scheme.
Contractions of supersymmetric extensions have usually been
considered separately, for the most relevant cases \cite{Hus,Zu}.
A BBL-classification however allows us to treat non-standard
models like Carroll and Newton algebras in unified manner. This
provides an alternative perspective to the numerous works
developed in connection to nonrelativistic limits of
supersymmetric theories \cite{Zu,flie}.

We briefly recall the notion of contraction. Given a Lie algebra
$\mathfrak{g}$ with structure tensor ${C_{ij}^{k}}$ over a fixed basis
$\left\{X_{i}\right\},i=1,..,n$, a linear redefinition of the
generators via a matrix $A\in GL(n,\mathbb{R})$ gives the
transformed structure tensor
\begin{equation}
{C^{\prime}}_{ij}^{
n}=A_{i}{}^{k}A_{j}{}^{\ell}(A^{-1})_{m}{}^{n}C_{k\ell}^{m}.
\label{BW}
\end{equation}

Considering a family $\Phi_{\epsilon}\in GL(n,\mathbb{R})$ of
non-singular linear maps of $\mathfrak{g}$, where $\epsilon\in (0,1]$,
for any $X,Y\in\mathfrak{g}$ we define
\begin{equation}
\left[X,Y\right]_{\Phi_{\epsilon}}:=\Phi_{\epsilon}^{-1}\left[\Phi_{\epsilon}(X),\Phi_{\epsilon}(Y)\right],
\end{equation}
which obviously reproduces the brackets of the Lie algebra over
the transformed basis. Actually this is nothing but equation
(\ref{BW}) for a special kind of transformations. Now suppose that
the limit
\begin{equation}
\left[X,Y\right]_{\infty}:=\lim_{\epsilon\rightarrow
0}\Phi_{\epsilon}^{-1}\left[\Phi_{\epsilon}(X),\Phi_{\epsilon}(Y)\right]
\label{Ko}
\end{equation}
exists for any $X,Y\in\mathfrak{g}$. Then equation (\ref{Ko}) defines
a Lie algebra $\mathfrak{g}^{\prime}$ which is a contraction of
$\mathfrak{g}$, since it corresponds to a limiting point of the orbit.
We say that the contraction is non-trivial if $\mathfrak{g}$ and
$\mathfrak{g}^{\prime}$ are non-isomorphic, and trivial otherwise. If
there is some basis $\left\{Y_{1},..,Y_{n}\right\}$ such that the
contraction matrix $A_{\Phi}$ adopts the form
\begin{equation}
(A_{\Phi})_{ij}= \delta_{ij}\epsilon^{n_{j}},\quad
n_{j}\in\mathbb{Z},\nonumber
\end{equation}
the contractions is called a generalized In\"on\"u-Wigner
contraction (gen. IW). In this sense contractions were originally
introduced in \cite{IW}, to describe continuous transitions from
relativistic to non-relativistic physics. It follows at once from
these considerations that contractions are transitive, i.e., if
$\mathfrak{g}$ contracts onto $\mathfrak{g}^{\prime}$ and
$\mathfrak{g}^{\prime}$ onto $\mathfrak{g}^{\prime\prime}$, then
$\mathfrak{g}$ contracts onto $\mathfrak{g}^{\prime\prime}$. This property
will be useful in the following. The concept of contraction can be
generalized without effort to other algebraic structures, in
particular non-associative algebras \cite{Ly}.

\medskip

In this work we focus primarily on generalized In\"on\"u-Wigner
(IW) contractions. As the generators of kinematical Lie algebras
are identified with physical operators, contractions obtained by
re-scaling certain of its generators still preserve this physical
meaning, up to some phase transitions for the rescaled elements.
Even if successive composition of IW-contractions is not
necessarily equivalent to a general IW-contraction, i.e., the
transitivity does not necessarily preserve diagonalization
properties, in each step we deal with diagonal transformations,
which enables us to interpret how the symmetry changes when
modifying the main parameters.

\section{Supersymmetric extensions of the de Sitter algebras}

According to the BLL classification, our starting point must be
the supersymmetric extension of the de Sitter algebras. The
difference between these is just the signature of the metric
tensor. We give the algebraic structure for the
$\mathfrak{so}(2,3)$ algebra. Consider the usual basis
$\left<L_{MN}=-L_{NM}, 0\le M < N \le 4\right>$ with commutation
relations
\begin{equation}\label{so23}
\left[L_{MN}, L_{PQ}\right]= \eta_{NP} L_{MQ} - \eta_{MP} L_{NQ} +
\eta_{QN} L_{PM} - \eta_{QM} L_{PN}, \end{equation} where
$\eta_{MN}=\text{diag}(1,-1,-1,-1,1)$. The generators $L_{MN}$
with $M,N\neq 4$ span the Lorentz algebra $\frak{so}(1,3)$. In the
basis $<L_{\mu \nu},P_\mu=L_{\mu 4}, \ \mu,\nu=0,\cdots,3>$ the
non-vanishing commutation relations are rewritten as

\begin{eqnarray}\label{so23L}
 \left[L_{\mu \nu}, L_{\rho \sigma}\right]&=&
\eta_{\nu \rho} L_{\mu \sigma} - \eta_{\mu \rho} L_{\nu \sigma} +
\eta_{\sigma \nu} L_{\rho \mu} - \eta_{\sigma \mu}
L_{\rho \nu},\nonumber\\
\left[L_{\mu \nu}, P_{\rho}\right]&=& \eta_{\nu \rho } P_\mu -
\eta_{\mu \rho}
P_\nu,\nonumber\\
\left[P_\mu, P_\nu\right]&=& L_{\nu \mu}. \end{eqnarray}

In terms of the BLL basis, with $K_{i}=L_{0i}$, $P_{i}=L_{i4}$,
$H=L_{04}$ and $L_{i}=L_{jk}$, where $i,j,k$ are taken in cyclic
order, the brackets are expressed as:
\begin{eqnarray}\label{so23bis}
\begin{array}{llll}
\left[L_{i}, L_{j}\right]= L_{k},& \left[L_{i}, K_{j}\right]=
K_{k},& \left[L_{i}, P_{j}\right]=
P_{k},& \left[K_{i}, K_{j}\right]= -L_{k},\nonumber \\
\left[P_{i}, P_{j}\right]= -L_{k},& \left[K_{i}, P_{j}\right]=
-\delta_{i j}H,& \left[K_{i}, H\right]=-P_{i},& \left[P_{i},
H\right]= K_{i}.
\end{array}
\end{eqnarray}
The remaining classical kinematical algebras and their commutators
are given in Table 1.

\begin{table}
\begin{indent}
\caption{{\small Non-vanishing brackets of classical kinematical
algebras in the standard basis. The common brackets to all Lie
algebras below are those corresponding to space isotropy:
$\left[\mathbf{L,L}\right]=\mathbf{L}$,
$\left[\mathbf{L,K}\right]=\mathbf{K}$ and
$\left[\mathbf{L,P}\right]=\mathbf{P}$.}}
\begin{tabular}
[c]{l|cccccccccc}\hline\hline & $\frak{so}\left(  2,3\right)  $ &
$\frak{so}\left(  1,4\right) $ & $I\frak{so}\left(  1,3\right)  $
& $I\frak{so}\left( 4\right) $ & $I\frak{so}\left(  1,3\right)
^{\prime}$ & Carroll & $Ne^{\exp}$ & $Ne^{osc}$ & $G\left(
2\right)  $ & $G\left( 2\right) ^{\prime}$\\\hline $\left[
\mathbf{K,K}\right]  $ & $-\mathbf{L}$ & $-\mathbf{L}$ &
$-\mathbf{L}$ & $0$ & $0$ & $0$ & $0$ & $0$ & $0$ & $0$\\
$\left[  \mathbf{K,P}\right]  $ & $-H$ & $-H$ & $-H$ & $-H$ & $-H$
& $-H$ &$0$ & $0$ & $0$ & $0$\\
$\left[  \mathbf{P,P}\right]  $ & $-\mathbf{L}$ & $\mathbf{L}$ &
$0$ &$\mathbf{L}$ & $-\mathbf{L}$ & $0$ & $0$ & $0$ & $0$ & $0$\\
$\left[  \mathbf{K,}H\right]  $ & $-\mathbf{P}$ & $-\mathbf{P}$ &
$-\mathbf{P}$ & $0$ & $0$ & $0$ & $-\mathbf{P}$ & $-\mathbf{P}$ &
$-\mathbf{P}$ & $0$\\
$\left[  \mathbf{P,}H\right]  $ & $\mathbf{K}$ & $-\mathbf{K}$ &
$0$ & $-\mathbf{K}$ & $\mathbf{K}$ & $0$ & $-\mathbf{K}$ &
$\mathbf{K}$ & $0$ & $-\mathbf{K}$\\\hline\hline
\end{tabular}
\end{indent}
\end{table}

Since Poincar\'e and Para-Poincar\'e ($I\frak{so}(1,3)^{\prime}$
in Table 1) algebras are isomorphic as Lie algebras, though
physically different, as happens with the two Galilei algebras, we
will only consider the standard Poincar\'e and Galilei algebras in
the following.

\subsection{The $\mathfrak{osp}(1|4)$ algebra}
We construct a supersymmetric extension of the anti-de Sitter
algebra starting from the real forms of
$\mathfrak{osp}(1|4,\mathbb C) =\mathfrak{sp}(4,\mathbb C) \oplus
\mathbb{C}^4$. Since only $\mathfrak{so}(2,3)$ admits a
four-dimensional real spinor  representation, only
$\mathfrak{so}(2,3)$ will have a supersymmetric extension.

Considering the decomposition $\mathfrak{osp}(1|4)=
\mathfrak{so}(2,3)\oplus\mathbb{R}^4 =\left<L_{\mu \nu},
P_\mu\right> \oplus \left< S_\alpha, \bar S^{\dot \alpha}\right>$,
where $\alpha, \dot \alpha =1,2, (S_\alpha, \bar S^{\dot \alpha})$
is a four dimensional Majorana spinor\footnote{$S^{\alpha}{}^\star
= \bar S_{\dot \alpha}$, $^\star$ denoting complex conjugation.}
of $\mathfrak{so}(1,3)$. Consider the Dirac $\Gamma-$ matrices
$\Gamma_{\mu}=\left(
\begin{array}
[c]{cc}%
0 & \sigma_{\mu}\\
\overline{\sigma}_{\mu} & 0
\end{array}
\right)  ,\;\Gamma_{4}=\left(
\begin{array}
[c]{cc}%
1 & 0\\
0 & -1
\end{array}
\right)$, where $\sigma_\mu=(1,\sigma_i), \bar \sigma_\mu =
(1,-\sigma_i)$ and $\sigma_i$ ($i=1,2,3$) denote the Pauli spin
matrices ($\sigma_\mu \to \sigma_\mu{}_{\alpha \dot \alpha},\;
\bar \sigma_\mu \to \bar \sigma_\mu{}^{\dot \alpha  \alpha}$).
Consider now $S_\alpha =\varepsilon_{\alpha\beta}S^\beta$,
$S^\alpha =\varepsilon^{\alpha\beta}S_\beta$, $\bar
S_{\dot\alpha}=\bar \varepsilon_{\dot\alpha \dot\beta}\bar
S^{\dot\beta}$, $\bar S^{\dot\alpha} =\bar
\varepsilon^{\dot\alpha\dot\beta}\bar S_{\dot\beta}$   with
$\varepsilon, \bar \varepsilon$ antisymmetric matrices given by
$\varepsilon_{12} = \bar \varepsilon_{\dot 1\dot 2}=-1$ and
$\varepsilon^{12} = \bar \varepsilon^{\dot 1\dot 2}=1$,
respectively. Defining the Majorana spinor $S_{A}=\left(
\begin{array}
[c]{c}%
S_{\alpha}\\
\bar S^{\dot \alpha}
\end{array}
\right)$ and introducing the anti de Sitter generators into the
spinor representation, we obtain the matrices
\begin{equation}
\Gamma_{\mu\nu}=\frac{1}{4}\left(
\begin{array}
[c]{cc}%
\left(  \sigma_{\mu}\overline{\sigma}_{\nu}-\sigma_{\nu}\overline{\sigma}%
_{\mu}\right)   & 0\\
0 & \left(  \overline{\sigma}_{\mu}\sigma_{\nu}-\overline{\sigma}_{\nu}%
\sigma_{\mu}\right)
\end{array}
\right)  ,\;\Gamma_{\mu4}=\frac{1}{2}\left(
\begin{array}
[c]{cc}%
0 & -\overline{\sigma}_{\mu}\\
\overline{\sigma}_{\mu} & 0
\end{array}
\right).
\end{equation}
In order to express the orthosymplectic algebra with real
structure constants, we need to consider generators in the
fermionic sector which respect to the following convention: we use
rescaled  $S_A$'s such that $S_\alpha^\star =i \bar S_{\dot
\alpha}$ and $\bar S_{\dot \alpha}^\star = i S_\alpha$. Then
$\mathfrak{osp}(1|4)$ reads \begin{equation} \left\{S_A,
S_B\right\} = b\ \Gamma_{MN}{}_A{}^D C_5{}_{BD} L^{MN}, b\neq
0,\nonumber
\end{equation}
where $C_5^{AB}=\left(
\begin{array}{cc}
\epsilon^{\alpha \beta}&0 \\
0&\bar  \epsilon_{\dot \alpha \dot \beta}
\end{array}
\right)$.\footnote{Since in our algebra we have $S^{*}_{\alpha}=i
\bar S_{\dot \alpha}$ there is a $e^{i \pi/4}$ factor for our
supercharge with respect to the usual conventions. As a
consequence, there is no $i$ factor in the bracket $\{S_\alpha,
\bar S_{\dot \alpha}\}$.} Choosing the normalization $b=4$ we
finally obtain the brackets
\begin{eqnarray}
\label{osp} &&\left[L_{\mu \nu}, S_\alpha\right]= (\sigma_{\mu
\nu})_\alpha{}^\beta S_\beta, \ \ \left[L_{\mu4} S_\alpha\right]=
-\frac12 \sigma_{\mu \alpha \dot \alpha} \bar S^{\dot \alpha},\ \
\left[L_{\mu \nu}, \bar S^{\dot \alpha}\right]= (\bar \sigma_{\mu
\nu})^{\dot \alpha}{}_{\dot \beta} \bar S^{\dot \beta}, \ \
\left[L_{\mu4}, \bar  S^{\dot \alpha}\right]= \frac12 \bar
\sigma_{\mu}^{\dot  \alpha  \alpha} S_{ \alpha},
\nonumber \\ \nonumber \\
&&\left\{S_\alpha, S_\beta\right\}= 4(\sigma^{\mu \nu})_{\alpha
\beta} L_{\mu \nu}, \ \left\{\bar S^{\dot \alpha}, \bar S^{\dot
\beta}\right\}= -4(\bar \sigma^{\mu \nu})^{\dot \alpha \dot \beta}
L_{\mu \nu},\ \left\{S_\alpha, \bar S_{\dot \beta}\right\}= 2
(\sigma^\mu)_\alpha{}_{\dot \beta} P_\mu. \nonumber
\end{eqnarray}
The brackets of the bosonic sector (\ref{so23L}) must be added to
the former. We observe that the convention adopted in this work is
different from the standard one \cite{wb,freu}.

\section{Supersymmetric Poincar\'e algebras}

The Poincar\'e algebra is classically derived from the anti-de
Sitter algebra by means of the contraction defined by the
transformations $L'_{\mu \nu}=L_{\mu \nu},\; P'_\mu =\varepsilon
P_\mu$ and taking the limit $\varepsilon \to 0$. The non-vanishing
brackets are
\begin{eqnarray}\label{poin} \left[L'_{\mu \nu}, L'_{\rho
\sigma}\right]&=& \eta_{\nu \rho } L'_{\mu \sigma} - \eta_{\mu
\rho } L'_{\nu \sigma}
+ \eta_{\sigma \nu } L'_{\rho \mu } - \eta_{\sigma \mu } L'_{\rho \nu}, \nonumber \\
\left[L'_{\mu \nu}, P'_\rho\right]&=&\eta_{\nu \rho} P'_\mu
-\eta_{\mu \rho} P'_\nu.
\end{eqnarray}
We remark that the isomorphism of the anti-de Sitter algebra with
$\frak{sp}(4,\mathbb{R})$ implies that the choice for the
contraction in the supersymmetric case must be the orthosymplectic
algebra $\mathfrak{osp}(1|4)$.\footnote{The ``natural" choice
$\mathfrak{osp}(5|N)$ violates the principle requiring that the
elements of the Fermi sector transform as Lorentz spinors.}
\cite{freu}. To the previous generators, we  add transformed generators
in the Fermi sector of $\mathfrak{osp}(1|4)$ defined by $Q_\alpha
= \varepsilon^a S_\alpha,\; \bar Q^{\dot \alpha} = \varepsilon^a
\bar S^{\dot \alpha}.$ In these conditions, the limit for
$\varepsilon\rightarrow 0$ exists and defines a superalgebra only
if $2 a \geq 1$ is satisfied. For $a=\frac12$ we recover the
algebra with non-trivial brackets
\begin{equation}\label{susy}
\left[L'_{\mu \nu} Q_\alpha\right]= (\sigma_{\mu
\nu})_\alpha{}^\beta Q_\beta,\; \left[L'_{\mu \nu} \bar Q^{\dot
\alpha}\right]= (\bar \sigma_{\mu \nu})^{\dot \alpha}{}_{\dot
\beta} \bar  Q^{\dot \beta},\; \left\{Q_\alpha, \bar Q_{\dot
\beta}\right\}= 2 \sigma^\mu{}_{\alpha \dot \beta } P'_\mu,
\end{equation}
which is the well-known supersymmetric algebra. This
supersymmetric extension of the Poincar\'e algebra was first
proposed in 1971, although its formal introduction in physics is
considered \cite{GL}. Since it arises as contraction of
$\mathfrak{osp}(1|4)$, we get a consistent contraction pattern for
supersymmetric extensions.

We remark the existence of another Poincar\'e-compatible
supersymmetric extension arising as contraction of
$\mathfrak{osp}(1|4)$ \cite{Ko}. However, no hermitian
representations for the generators in the Fermi part exist, thus
the model is of no use in supersymmetric considerations.

\section{Supersymmetric Galilei algebra}

Contracting the Poincar\'e algebra $I\frak{so}(1,3)$ using the
transformations \begin{equation}\label{IW-Gal} L'_i=L_i, \ \
K'_i=\varepsilon K_i, \ \ P'_i= \varepsilon P_i, \ \ H'=H.
\end{equation}
(see Table 1). A supersymmetric extension is obtained by adding
the following odd generators to the previous: \begin{equation}
\label{IW-supergal} Q'_\alpha = \epsilon^a Q_\alpha, \ \ \bar
Q'_\alpha = \epsilon^b \bar Q_\alpha. \end{equation}

Taking into account the embedding $\mathfrak{so}(3) \subset
\mathfrak{so}(1,3)$, it turns out that both representations
$\left< Q_\alpha\right>$ and $\left< \bar Q_{\dot \alpha}\right>$
are equivalent, although complex conjugate. We denote $\bar
Q_{\dot \alpha} \to \bar Q_{\alpha}$ and $\sigma_i{}_{\alpha \dot
\beta} \to \sigma_i{}_{\alpha \beta}$. The contraction of the
supersymmetric Poincar\'e algebra with respect to these
transformations lead to a superalgebra whenever the condition $a +
b -1 \ge 0$ is satisfied. Among these supersymmetric extensions,
that given by $a+b-1=0$ is the usual $N=2$ supersymmetric
extension (without central charge) of the Galilei algebra:

\begin{eqnarray}\label{sugal}
\begin{array}{cc}
\begin{array}{cc}
\left[L'_k, Q'_\alpha\right]= -\frac{i}{2}
(\sigma_k)_\alpha{}^\beta Q'_\beta,& \left[K'_k,
Q'_\alpha\right]=0, \\
\left[L'_k, \bar Q'_\alpha\right]= -\frac{i}{2}
(\sigma_k)_\alpha{}^\beta \bar Q'_\beta,& \left[K'_k, \bar
Q'_\alpha\right]=0,
\end{array}
& \left\{Q'_\alpha, \bar Q'_\beta\right\}= 2 \sigma_{i}{}_{\alpha
\beta} P^{\prime i},
\end{array}
\end{eqnarray}

Using the transitivity of contractions, it can be easily seen that
the extended Galilei algebra can be obtained contracting the
supersymmetric extension of the anti de Sitter algebra. We remark
that other extension considered in the literature is obtained
using a special $\mathbb{Z}_{4}$-grading of the Poincar\'e
superalgebra \cite{Hus}. It seems however that no supersymmetric
field models based on this Galilei superalgebra have been
developed.

\section{Supersymmetric Carroll algebra}

The Carroll algebra, a quite strange object, appears as an
alternative non-relativistic limit of the Poincar\'e group, and
was first described in the work \cite{LL,LL1}. It is obtained from
the Poincar\'e algebra through the contraction determined by
rescaling the boosts and time translations:
\begin{equation}\label{IW-C}
 J'_i=J_i, \ \ K'_i = \epsilon K_i, \ \ P'_i =
P_i, \ \ H'= \epsilon H.
\end{equation}
Although the Carroll algebra has played no distinguished role in
kinematics, it appears in the study of tachyon condensates in
string theory \cite{Gib}. In this context a possible
interpretation of this limit and supersymmetric extensions regain
 some interest. To construct
a supersymmetric extension of the Carroll algebra, we add to the
generators specified in (\ref{IW-C}) the additional generators of
the symmetric part
\begin{eqnarray}\label{supergal}
 Q'_\alpha = \epsilon^a Q_\alpha, \ \ \bar
Q'_\alpha = \epsilon^b \bar Q_\alpha.
\end{eqnarray}

We therefore obtain a contraction of $\mathfrak{osp}(1|4)$ if the
constraint $a+b-1 \ge 0$ is satisfied. Again, for the special case
$a+b-1=0$, we get a $N=2$ supersymmetric extension of the Carroll
algebra given by
\begin{eqnarray}\label{suoer-Car}
\begin{array}{cc}
\begin{array}{cc}
\left[L'_k, Q'_\alpha\right]= -\frac{i}{2}
(\sigma_k)_\alpha{}^\beta Q'_\beta,& \left[K'_k,
Q'_\alpha\right]=0, \\
\left[L'_k, \bar Q'_\alpha\right]= -\frac{i}{2}
(\sigma_k)_\alpha{}^\beta \bar Q'_\beta,& \left[K'_k, \bar
Q'_\alpha\right]=0,
\end{array}
& \left\{Q'_\alpha, \bar Q'_\beta\right\}= 2 \delta_{\alpha \beta}
H.
\end{array}
\end{eqnarray}

For this supersymmetric extension we find an interesting feature,
namely, that in the fermionic  sector the symmetric product
reproduces a structure of Clifford algebra \cite{clif}. It is not
difficult to verify that this algebra corresponds to the $N=4$
supersymmetric quantum mechanics extension of the Galilei algebra.
The main difference between this and the Galilean supersymmetric
model reside in the fact that the supercharges are in the spinor
representation of the rotation group.

\section{Supersymmetric Newton algebra}

Newton-Hooke kinematical algebras arise as contractions of the de
Sitter algebras, and there is no relation, by means of continuous
contractions, i.e., a limiting process, between these and the
Poincar\'e algebra. However, in analogy with the latter, they have
the Galilei algebra as non-relativistic limit. The contraction is
determined by the transformations \begin{equation}\label{IW-N}
P'_i=\epsilon P_i, \ \ K'_i=\epsilon K_i. \end{equation}

The limit $\epsilon \to 0$ corresponds to the oscillating Newton
Lie algebra. As observed above, properties of these models are
close, in some sense, to those of the de Sitter algebras, like
curvature properties of space-time. In a different context, they
have been studied to some extent in non-relativistic brane theory
\cite{Her}. A supersymmetric extension of the Newton algebra is
obtained by joining to the generators of (\ref{IW-N}) the
additional elements
\begin{equation}
\label{IW-sN} Q_\alpha= \epsilon^a S_\alpha, \ \ \bar Q_\alpha=
\epsilon^b \bar S_\alpha.
\end{equation}

If we compute the corresponding brackets, it follows that the
action of $P'$ on $Q$ and $\bar Q$ is trivial, although the
anti-commutators lead to the constraint $a+b-1 \ge 0$. Here there
is only one possibility to obtain a contraction with non-trivial
symmetric product, given by the condition $a+b-1=0$. Therefore the
$N=2$ supersymmetric extension has brackets
\begin{eqnarray}
\begin{array}{cc}
\begin{array}{cc}
\left[L'_k, Q'_\alpha\right]= -\frac{i}{2}
(\sigma_k)_\alpha{}^\beta Q'_\beta,& \left[K'_k,
Q'_\alpha\right]=0, \\
\left[L'_k, \bar Q'_\alpha\right]= -\frac{i}{2}
(\sigma_k)_\alpha{}^\beta \bar Q'_\beta,& \left[K'_k, \bar
Q'_\alpha\right]=0,
\end{array}
& \left\{Q'_\alpha, \bar Q'_\beta\right\}= 2 \sigma_{i}{}_{\alpha
\beta} P^{\prime i},
\end{array}
\end{eqnarray}

Using a different approach, an extension equivalent to this one
was already considered in \cite{Hus}. In particular, it is
straightforward to verify that this supersymmetric Newton
extension contracts onto the supersymmetric Galilei algebra
(\ref{sugal}).

\section{Supersymmetric static algebra}

A supersymmetric extension of the static Lie algebra appears as
contraction of any superextension, by simply considering the pure
inequalities in all the constraints obtained previously. Actually
this algebra is nothing but the sum of the classical static
algebra and a four dimensional space with trivial symmetric
product. In this sense, this extension can be seen as the final
superalgebra that still preserves the condition of space isotropy.
The general contraction pattern can be resumed in the following
diagram:

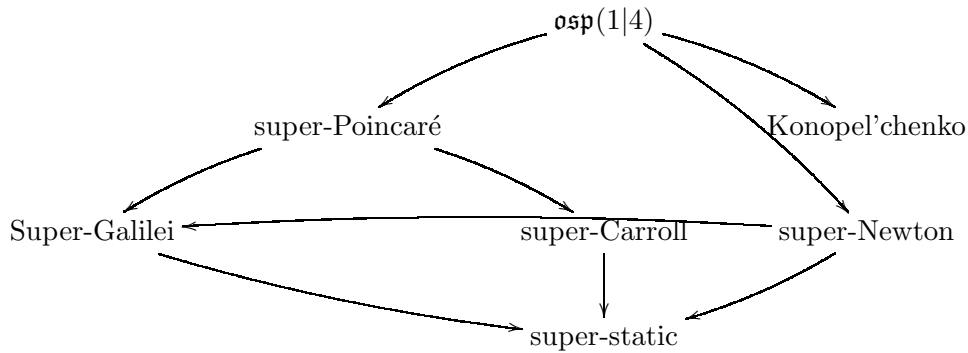
\begin{figure}
\caption{Contraction hierarchy of supersymmetric extensions}
$$
\xymatrix{
&& \mathfrak{osp}(1|4)\ar@/_/[dl] \ar@/^/[dr] \ar@/^/[ddr] &&  \\
&\text{super-Poincar\'e}   \ar@/_/[dl] \ar@/^/[dr]
&&\text{Konopel'chenko} \\
\text{Super-Galilei} \ar@/_/[drr]&& \text{super-Carroll}\ar@//[d]&
\text{super-Newton}\ar@/^/[dl] 
 \ar@/_/[lll] \\
&&\text{super-static} }
$$

\end{figure}

\section{Final remarks}

 Applying the contraction ansatz of Bacry and L\'evy-Leblond to
classify the possible kinematics, we have extended the method, to
obtain a similar classification of ``kinematical" supersymmetric
extensions of these Lie algebras. Various possibilities arise,
although only those leading to physically interesting
superalgebras, due to the non-apparent relation with field
theoretic realizations for the remaining solutions. We however
remark that this general approach, leading to all possible
supersymmetric extensions of kinematical algebras, including
exotic models like the alternative Poincar\'e or Galilean
superalgebras and their respective contractions, can be done in
analogy with the general analysis of \cite{Nu}. Work in this
direction is in progress.

We finally remark that the classical approach of \cite{bl} is
based on the assumption that parity and time reversal (PT) are
automorphisms of the kinematical group. This is well known to fail
for weak interactions, thus a relaxation of the hypothesis arising
from this general analysis could lead to physically interesting
models. In this sense, a kinematical classification of cubic
extensions has been worked out \cite{C77}, allowing to recover the
known extensions studied in the literature
\cite{flie,csusy,1dfsusy}. In this third step, the possibility of
discrete symmetries in the form of graded contractions must be
taken into account.

\subsection*{Acknowledgement}
One of the authors (RCS) acknowledges partial financial support
from the research projects MTM2006-09152 of the Ministerio de
Educaci\'on y Ciencia and CCG07-UCM/ESP-2922 of the U.C.M.-C.A.M.

\end{document}